\documentclass[aps,pre,preprint]{revtex4-2}
\usepackage{graphicx}%
\usepackage{amsmath,amssymb,amsfonts,bm}%
\newcommand{\vR}{\mathbf{R}}

\DeclareMathOperator\erfi{\text{erfi}}

\begin{document}

\title{Length--dependent residence time of contacts in simple polymeric models}

\author{Edoardo Marchi}
\affiliation{Department of Physics, Universit\`a degli Studi di Milano  and INFN, via Celoria 16, 20133 Milano, Italy}
\author{Guido Tiana}
\affiliation{Department of Physics and Center for Complexity and Biosystems, Universit\`a degli Studi di Milano and INFN, via Celoria 16, 20133 Milano, Italy}

\email{guido.tiana@unimi.it}
\date{\today}

\begin{abstract}
Starting from the reported experimental evidence that the residence time of contacts between the ends of biopolymers is length dependent, we investigate the kinetics of contact breaking in simple polymer models from a theoretical point of view. We solved Kramers equation first for an ideal chain and then for a polymer with attracting ends, and compared the predictions with the results of molecular dynamics simulations. We found that the mean residence time always shows a power--law dependence on the length of the polymer with exponent $-1$, although is significantly smaller when obtained from the analysis of a single trajectory than when calculated from independent initial conformations. Only when the interaction is strong ($\gg kT$) and the interaction range is small (of the order of the distance between consecutive monomers) does the residence time converge to that of the Arrhenius equation, independent of the length. We are able to provide expressions of the mean residence time for cases when the exact definition of contact is not available {\it a priori}, expressions that can be useful in typical cases of microscopy experiments.
\end{abstract}

\maketitle

\section{Introduction}

The encounter time between the ends of polymers has been extensively studied in the literature, both experimentally and theoretically. Fluorescence energy transfer experiments have shown that the distribution of encounter times in different types of polymers does not follow a thermally activated two--state scheme \cite{Bonnet1998,Wallace2001Non-ArrheniusHairpin}. The correct description of the encounter times requires the modelling of the diffusion of the polymer ends \cite{Szabo1980}. As a consequence, the mean first encounter time $\overline{\tau_b}$ depends on the length $\ell$ of the polymer separating the two ends. Such a dependence can be studied analytically for an ideal chain solving the Fokker-Planck equation under some approximations and gives $\overline{\tau_b}\sim\ell^{3/2}$ for long chains and $\overline{\tau_b}\sim\ell^2$ for short chains \cite{Amitai2012}. 

Less information is available for the residence time $\tau$ of polymeric contacts. As a first approximation, the opening of a contact could be described as a two--state Poisson process, independent of the length of the polymer \cite{Tiana2001}. However, indirect estimation of the average $\overline{\tau}$, obtained by comparing closing rates with the equilibrium stability of the contact in DNA and RNA hairpins, suggests a non--negligible dependence on $\ell$ \cite{Shen2001,Kuznetsov2008}. Also in the case of proteins, the unfolding rate was observed to depend on the length of the chain segment between native contacts; increasing this length by engineering aminoacidic linkers between native contacts \cite{Nagi1999} or by creating circular permutants of a protein \cite{Viguera1997}, the unfolding rate is affected.

The mean residence time (or mean opening time) is quite relevant in several biophysical scenarios where a minimum time is required to complete a process. One example is the increasing importance of the dissociation rate $k_{\text{off}}$ in drug design \cite{Copeland2006}. In fact, in {\it in vivo} scenarios where the concentration of ligands close to the target may be highly fluctuating, the association and dissociation rate are critical to determine the efficiency of the drug.  The dissociation rate has been proposed to be a better determinant of drug efficacy {\it in vivo}  than the equilibrium constant \cite{Pan2013}. Another example is the activation of genetic transcription, which occurs in bursts and it is controlled by the formation of loops  between enhancers and promoters in chromatin \cite{Amitai2017}. The loops should last long enough for multiple protein complexes to be recruited \cite{Yang2024EnhancerActivation}. In this case, contacts are stabilised by complex out--of--equilibrium mechanisms involving energy-consuming proteins \cite{Fudenberg2016a}. The duration of the bursts is in general much longer than then the expected residence time of the loops; this difference requires a deeper understanding of the molecular mechanisms involved, including those associated with the polymeric nature of the system \cite{Lammers2020}. For proteins, the residence time of native contacts is important because it has been shown that the variation in protein stability upon mutation is mainly determined by this quantity, rather than by changes in folding rate \cite{Shah2017}.

We have studied the duration of contacts in simpler polymeric systems at equilibrium, which, despite their simplicity, can be used as a reference point for more sophisticated and realistic analyses. First, we studied the residence time of contacts in an ideal chain, where the definition of what constitutes a contact is purely arbitrary. We defined the two ends of the polymer as forming a contact if their mutual distance is less than a threshold $\epsilon$, and studied the dependence of the residence time as a function of this parameter, which is not naturally suggested by the system. We derived the equations that predict the residence time of the contact as a function of the parameters that define the system and compare their prediction with molecular dynamics simulations. Even in this simple case, the residence time is dependent on the length of the polymer.

Then, we extended the study to that of an ideal chain whose ends attract each other with an energy described by a spherical well of width $R_p$. In general, the value of $R_p$ for a real system is difficult to know and in experiments the definition of a clear bound that defines the formation of a contact is arbitrary. In the absence of a definition {\it a priori} of the contact based on the molecular properties of the system, one can use algorithms that derive a definition from the data, such as hidden Markov models \cite{Mach2022}. 

The model we have developed allows us to study the residence time without knowing the contact length in advance. It describes not only the case where the contact is defined to match the width of the potential well, but also the cases where they are different. In this way, the true contact length and the interaction energy of the contact can be derived from the data reporting the mean residence time at different distances.

We have found that the dependence of the mean residence time on the length $\ell$ of the polymer depends on the balance between two factors. The first is the residence time, defined from a single initial conformation, which has a dependence on $\ell$ caused by the diffusion of the chain ends within the contact length. The second is the averaging over all initial conditions to give the mean residence time. We have studied two different experimentally relevant initial conditions, both of which depend on $\ell$. The resulting mean residence time thus has a non--trivial dependence on the length of the polymer.

\section{Residence time of the ideal chain in simulations is length--dependent} \label{sect:sim}

We have performed Langevin dynamics simulations of an ideal chain of $10^3$ monomers (details in Appendix A) with LAMMPS \cite{Thompson22}. A contact between two monomers $i$ and $j$ with separation $\ell=|i-j|$ along the chain is defined if the Euclidean distance $R$ between them is less than a threshold $\epsilon$. The residence time of a contact is defined as the time taken by the distance to reach $\epsilon$ for the first time.

The mean residence time of a contact depends on the initial conditions of the system. We tested two cases of practical interest. In the former (initial condition A), we sampled random equilibrium conformations of the ideal chain and selected a set of them that are distant in time and have a contact as initial conformations for the dynamics. In this case, initial conditions are uncorrelated with each other. This corresponds to the experimental situation in which multiple polymers are tracked simultaneously until each contact that was formed is broken (e.g., chromatin loci in different nuclei by fluorescence microscopy). In another experimental scenario, the formation and disruption of a contact is tracked in the same polymer along a single trajectory. We then ran a long simulation of a polymer, using the conformation recorded just after the events of contact formation as starting points to calculate the residence time and collecting statistics (initial condition B). Operatively, one analyzes the timeframes of a trajectory recorded at intervals $\Delta t$ and calculates the (discretized) time span between the frame in which the contact is formed after a frame in which it is not formed, and when it is broken.  In this case, the initial conformations are biased by the condition of being close to the formation event of the contact and may be correlated with each other. This non--equilibrium distribution depends on the resolution $\Delta t$ used to monitor the trajectory, becoming equal to those of initial conditions A for large $\Delta t$. Nevertheless, this strategy is often used in experimental setups for operational convenience.

\begin{figure}
    \centering
    \includegraphics[width=\linewidth]{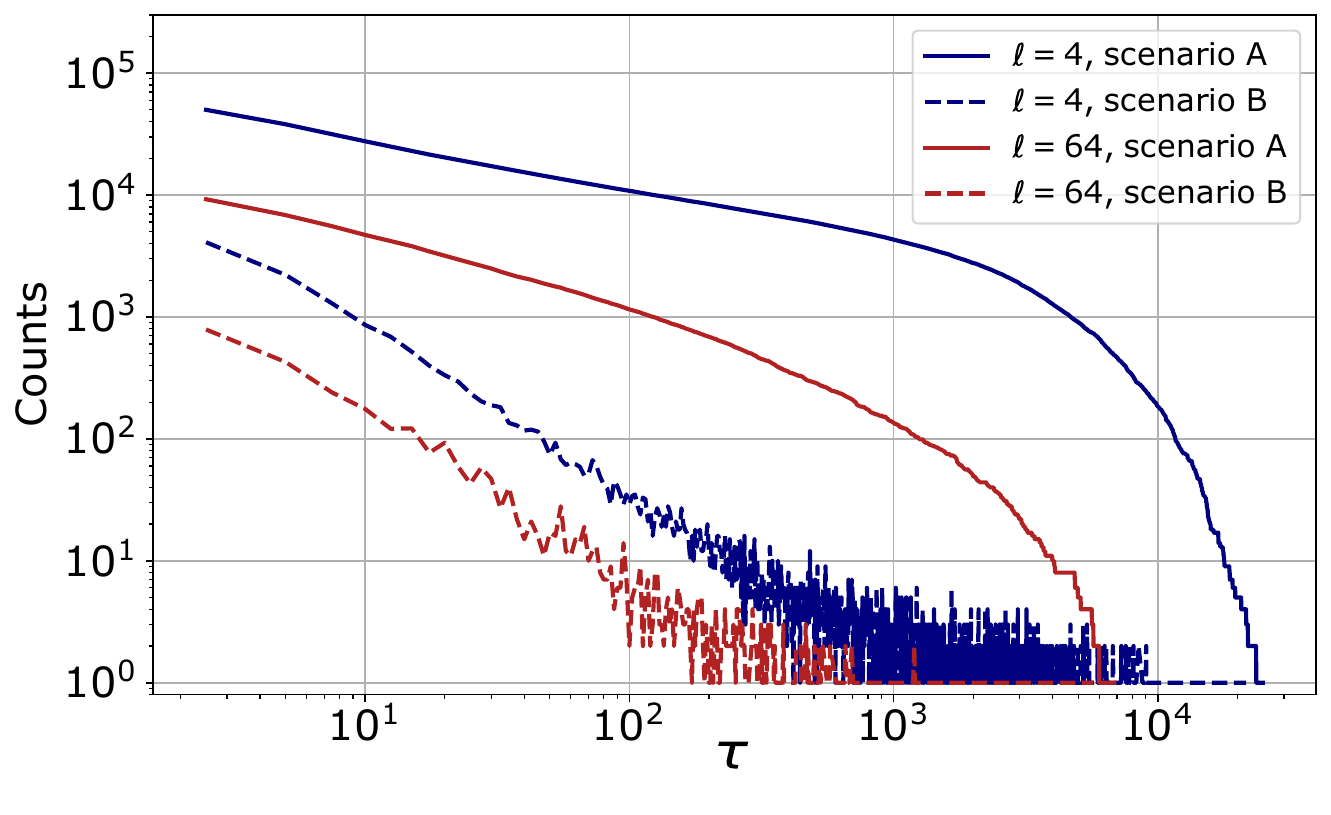}
    \caption{ The distribution of residence times simulated for $\ell=4$ and 64 for initial conditions A and B in log--log scale.}
    \label{fig:distr_times_ideal}
\end{figure}

The distribution of residence times (Fig. \ref{fig:distr_times_ideal}) depends on $\ell$ for both initial conditions A and B. Typically, it has a power--law  distribution $t^{-\alpha}$ with small exponent ($\alpha\lesssim 1$) at short times and an exponential cutoff, making the mean residence time finite. The residence times are shorter for initial conditions B because these are biased to be on the verge of contact break (Fig. S1 in the Supp. Mat.).  

\section{Mean residence time in the ideal chain} \label{sect:ideal}

The starting point of our treatment is a dimensional reduction from the space of polymeric conformations to the space of distances $R$ between the ends of the polymer. Assuming that the dynamical mode corresponding to $R$ is the slowest among  those associated with the other degrees of freedom of the chain, then $R$ undergoes a Markovian diffusive dynamics controlled by an effective force, which can be obtained from the gradient of the free energy \cite{Zwanzig1961}. Since the dynamics of $R$ corresponds to the lowest Rouse modes (of longest wavelength) and the time scales associated with these modes scale as the inverse square mode index, we expect that this assumption holds (and this even more true if there is, as in the case of Sect. \ref{sect:attractive}, an energy well that slows down the dynamics of $R$). In this case, the effective energy that controls $R$ arises just from the entropy of the chain,
\begin{equation}
    U(R)=kT\left[ \frac{3R^2}{2\ell a^2}-\log R^2 \right],
\end{equation}
where $\ell$ is the length of the chain and $a$ is the separation between consecutive monomers (cf. Appendix B). Moreover, we focus our attention on time scales that are much longer than the damping time due to friction (see Appendix \ref{sect:details}), so that inertial effects can be neglected and we can characterize the polymer with only its spatial coordinates.

The mean residence time for a contact between two monomers of a polymer undergoing Brownian dynamics and starting at a given distance $R<\epsilon$, can be obtained from the solution of Kramers equation for a diffusing system in the one--dimensional radial coordinate 
\begin{equation}
    \tau(R)=\frac{1}{D}\int_R^\epsilon dR'\; e^{\beta U(R')}\int_0^{R'}dR''\; e^{-\beta U(R'')},
    \label{eq:kramers}
\end{equation}
where $D$ is the diffusion coefficient, $\beta=1/kT$ is the inverse thermal energy and $U(R)$ is the effective energy acting between the monomers. 

The integrals in Eq. (\ref{eq:kramers}) can be solved exactly (cf. Appendix \ref{sect:residence_supp}) and give
\begin{equation}
    \tau(R)=\frac{1}{6D}\left[\epsilon^2\; _2F_2\left( \frac{3\epsilon^2}{2a^2\ell} \right)-R^2\; _2F_2\left( \frac{3R^2}{2a^2\ell} \right)  \right],
    \label{eq:tauR}
\end{equation}
where $_2F_2(x)$ is an abbreviation for the generalized hypergeometric function 
\begin{equation}
    _2F_2(x)\equiv\, _2F_2\left(
    \begin{array}{cc}
    1 & 1 \\ 2 & 5/2
    \end{array}, x
    \right).
\end{equation}

The mean residence time depend on the distribution $p_0(R)$ of initial conditions,
\begin{equation}
    \overline{\tau}=\frac{1}{6D}\,_2F_2\left( \frac{3\epsilon^2}{2a^2\ell} \right)-\frac{1}{6D}\int_0^\infty dR\,p_0(R)\, _2F_2\left( \frac{3R^2}{2a^2\ell} \right).
\end{equation}
Because hypergometric functions are difficult to handle, some approximations are necessary.

\subsection{Initial conditions of type A}

\begin{figure}
    \includegraphics[width=\linewidth]{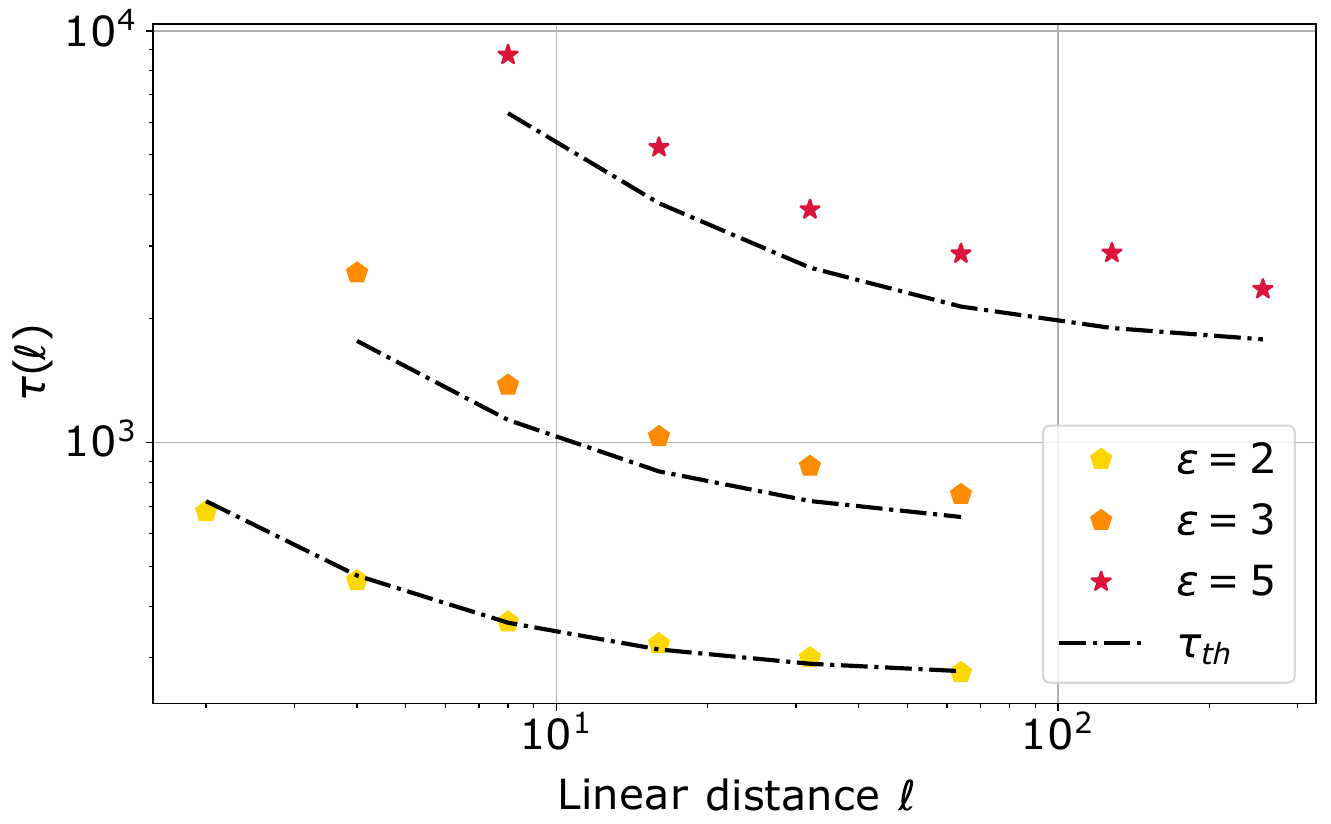}
    \caption{Mean residence time of contacts in the ideal chain as a function of   $\ell$ for different contact definitions $\epsilon$ for the initial conditions of type A, calculated from simulations (colored symbols) and predicted by Eq. (\protect\ref{eq:tau_th_A}) (dot--dashed lines).}
     \label{fig:tau_ideal_A}
\end{figure}

For initial conditions of type A, the equilibrium distribution of the ideal chain [Eq. (\ref{eq:eq_ideal}))] can be used as $p_0(R)$. If the length defining the contact is smaller than the equilibrium distance between the monomers ($\epsilon\leq a\ell^{1/2}$), the entropic force tends to break the contact and the mean residence time has an expression [Eq. (\ref{eq:tau_th_A})] that is in good agreement with the simulations (Fig. \ref{fig:tau_ideal_A}). The agreement is better for small $\epsilon$, corresponding to the approximation made to obtain Eq. (\ref{eq:tau_th_A}), and worsens as $\epsilon$ increases, showing a small downscale with respect to the simulated points.

For $a\ell^{1/2}$ of the order of $\epsilon$, the predicted expression for $\overline{\tau}$  can be approximated as [Eq. (\ref{eq:tauA_apf})]
\begin{equation}
    \overline{\tau}\approx \frac{0.17\epsilon^2}{D}-\frac{0.06 \,\epsilon\,a}{D} \sqrt{\ell},
    \label{eq:tau_icA1}
\end{equation}
and it decreases with $\ell$ until for small $\epsilon/a\ell^{1/2}$ satisfies
\begin{equation}
    \overline{\tau}=\frac{\epsilon^2}{15D}\left( 1 + \frac{48}{75}\frac{\epsilon^2}{a^2\ell} \right).
    \label{eq:tau_icA2}
\end{equation}

The simulated $\overline{\tau}$ shows a weak dependence on $\ell$ for $\epsilon<a\ell^{1/2}$ (Fig. \ref{fig:tau_ideal_A}), always remaining within the same order of magnitude. In fact, $\overline{\tau}$ is convex and the steepest part of the curve is that for which $\epsilon\approx a\ell^{1/2}$ [Eq. (\ref{eq:tau_icA1})], where its slope remains small when $0.06a\ll 0.17\epsilon$, that is when the length $\epsilon$ of the contact is greater than the inter--monomer length $a$.

For large $\ell$, the mean residence time saturates at $\epsilon^2/15D$ because the dependence on $\ell$ coming from $\tau(R)$ and from $p_0(R)$ cancels out. The decay is quite long range, scaling as $\ell^{-1}$. 

The dependence of the mean residence time on $\epsilon$ is quite strong, showing a quadratic behaviour [Eq. (\ref{eq:tau_icA1})] for the smaller $\ell$.

\subsection{Initial conditions of type B}

The initial conditions of type B are defined by the conditional probability that a contact will be formed at a given time, from which we start counting $\tau(R)$, provided that it has not been formed at the previous time during the observation of the contact. If the system is observed with time resolution $\Delta t$, then the distribution of initial distances can be obtained from the Chapman--Kolmogorov equation
\begin{equation}
    p_0(R)=\int_0^\infty dR'\, p_{\Delta t}(R,t+\Delta t|R',t)p(R',t|R'>\epsilon,t),
    \label{eq:chapman-kolmogorov}
\end{equation}
where the second factor in the integral is the probability that the starting position of the jump is $R'$, with no contact. It is calculated from the stationary distribution of an ideal chain, assuming $\ell>(\epsilon/a)^2$. The first term is the probability of making a jump from $R'$ to $R$. It is calculated in the Onsager-Machlup approximation, assuming that $\Delta t$ is small enough that the force does not change appreciably. We obtain (cf. Appendix \ref{sect:residence_supp})
\begin{equation}
\label{eq:p_0_B_approx}
    p_0(R)\approx \frac{\sqrt{\pi}}{\sqrt{4D\Delta t}} \left[1-\text{erf}\left(\frac{\epsilon-R}{(4D\Delta t)^{1/2}} \right) \right],
\end{equation}
which is independent on $\ell$ but has a strong dependence on $\Delta t$ (cf. Fig. S2 in the Supp. Mat.). The mean residence time can be approximated [Eq. (\ref{eq:supp_tau_final})] for $\ell\gg\epsilon^2/a^2$ as  
\begin{equation}
    \overline{\tau}\approx \frac{\ell a^2\sqrt{\pi\Delta t}}{12\epsilon^2\sqrt{D}}
    \left[ 
    \sqrt{\frac{2\pi a^2\ell}{3}} e^{\frac{3\epsilon^2}{2a^2\ell}} \text{erf}\left(\frac{\sqrt{3}\epsilon}{\sqrt{2a^2\ell}}\right)-2\epsilon
    \right].
    \label{eq:tauB2}
\end{equation}

\begin{figure}
    \includegraphics[width=\linewidth]{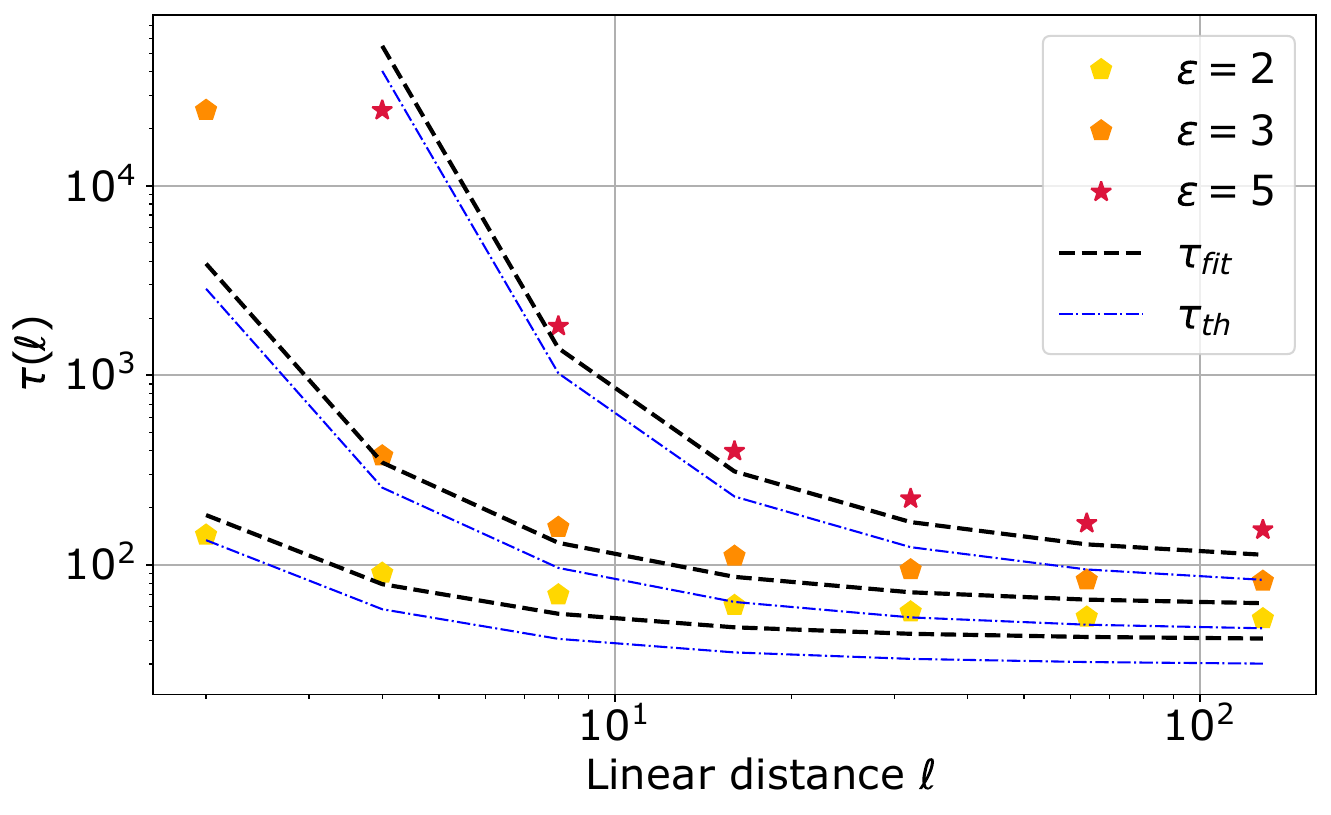}
    \caption{Mean residence time of contacts in the ideal chain as a function of $\ell$ for different $\epsilon$, calculated from simulations (colored symbols) for initial condition B ($\Delta t=2.5$). The dot--dashed blue lines are the predictions of the model using the nominal parameters of the model, while the dashed black lines are obtained fitting $p_0(R)$ from the simulations.}
     \label{fig:tau_ideal_B}
\end{figure}

The prediction of Eq.  (\ref{eq:tauB2}) is in qualitative agreement with the simulated points (blue dot--dashed curves in Fig. \ref{fig:tau_ideal_B}). However, the prediction seems to be scaled down from the data by a factor of $\approx 2$. The reason is that the distribution $p_0(R)$ of initial distances predicted by Eq. \ref{eq:p_0_B_approx} is slightly different from the simulated data (Fig. S3 in the Supp. Mat.) and the resulting $\overline{\tau}$ is strongly dependent on $p_0(R)$. Fitting the parameters of $p_0(R)$ from the simulation (Fig. S4 in the Supp. Mat.) instead of using the nominal values of $\epsilon$ and $D\Delta t$ leads to a better match of $\overline{\tau}$ (black dashed curves in Fig. \ref{fig:tau_ideal_B}).

The variability of $\overline{\tau}$ is much higher here than with initial conditions of type A (Fig. \ref{fig:tau_ideal_B}), mainly because of the exponential term in Eq. (\ref{eq:tauB2}). Moreover, the predictions become unreliable for the leftmost points, where $\epsilon\gtrsim a\ell^{1/2}$, because the main approximation we did [$\epsilon\ll a\ell^{1/2}$ in Eq. (\ref{eq:tau_th_A})] fails.

In the limit of large $\ell$, the mean residence time scales as
\begin{equation}
    \overline{\tau}\approx \frac{\sqrt{\pi \Delta t}\epsilon}{6\sqrt{D}}\left(1 + \frac{3}{5}\frac{\epsilon^2}{a^2\ell}\right).
\end{equation}
which reaches an asymptotic value different from that of initial condition A (Eq. \ref{eq:tau_th_A}) but at similar values of $\ell$.

The residence time depends on $\Delta t$ (Fig. S5 in the Supp. Mat.) because for initial conditions B the initial distribution $p_0(R)$ depends on this quantity (Fig. S2). For large $\Delta t$, the starting conformation to calculate the residence time becomes independent of the instant of contact formation and then $p_0(R)$ converges to that of initial conditions A. However in this limit the contact can break and re--form within the non--observed time $\Delta t$, making the calculation of the residence time imprecise. Consequently we wish to choose $\Delta t$ as small as possible.
Note that it is not possible to evaluate the mean residence time in the limit of $\Delta t\to 0$ because the Kramers solution from which we started the whole derivation [Eq. (\ref{eq:kramers})] is valid only in the overdamped regime $\Delta t\gg m/\gamma$, where $m$ is the mass of the moving monomer and $\gamma$ its friction coefficient. To get a reasonable result for small times, one should consider the inertia of the system, which would prevent the system from breaking the contact immediately after it is formed.

\section{Ideal chain with attracting ends} \label{sect:attractive}

The above treatment can be extended to the case of an ideal polymer whose ends interact with a spherical--well attracting potential $V(R)$ of width $R_p$ and depth $\beta E_w$. When $\epsilon<R_p$, the presence of the well has no effect on the mean residence time for any choice of the initial conditions, and thus the expressions found above are still valid. 

When $\epsilon> R_p$, the residence time $\tau(R)$ starting from a given point $R$ can be found integrating Eq. (\ref{eq:kramers}), exploiting the fact that the potential $V(R)$ is defined piece--wise. The result [Eq. (\ref{eq:tau01})] can be written in the form
\begin{equation}
    \tau(R)=\tau_0(R)+\tau_1(R),
    \label{eq:tauR_int}
\end{equation}
where $\tau_0(R)$ is that obtained for the ideal chain [Eq. (\ref{eq:tauR})], while $\tau_1(R)$ depends on $V(R)$ and has two different expressions depending on whether $R$ is greater or less than $R_p$ [Eq. (\ref{eq:tau1Rcases})]. A particularly simple expression can be found for large $\ell$, for which
\begin{equation}
    \tau(R)=\tau_0(R)+\frac{R_p^3}{2D}(e^{\beta E_w}-1)\left(\frac{1}{\max[R,R_p]}-\frac{1}{\epsilon}\right).
\end{equation}

A mean residence time can be found using the expression valid for any value of $\ell$ [Eq. (\ref{eq:tau01})] for the two choices of the initial conditions.

\subsection{Initial conditions of type A}

In this case, the initial probability $p_0(R)$ is Boltzmann distribution, controlled by the entropic term of the polymer and the attractive energy $V(R)$, 
\begin{equation}
    p_0(R) = \begin{cases} \frac{R^2}{Z} e^{-\frac{3R^2}{2\ell a^2}+\beta E_w} & \text{if } R < R_p \\
    \frac{R^2}{Z} e^{-\frac{3R^2}{2\ell a^2}} & \text{if } R \geq R_p,
    \end{cases}
\end{equation}
where the $Z$ is the normalization factor [Eq. (\ref{eq:zAint})]. The expression of $\tau(R)$ [Eq. (\ref{eq:tauR_int})] can be integrated together with $p_0(R)$, obtaining a non--trivial expression made of four terms [Eq. (\ref{eq:taumA4})].

\begin{figure}
    \includegraphics[width=\linewidth]{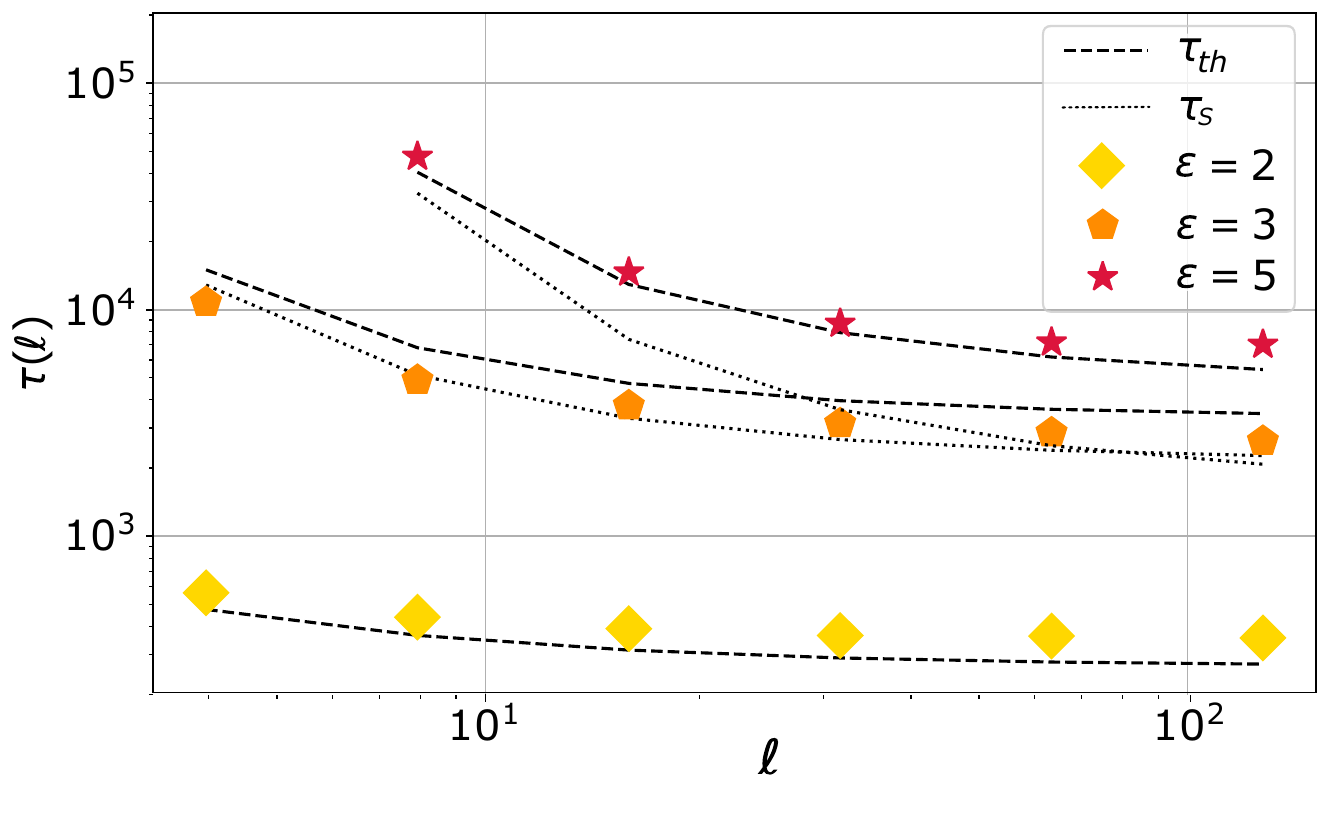}
    \caption{Mean residence time of contacts in the chain with attracting ends as a function of the linear separation $\ell$ for different definitions $\epsilon$ of the contact, predicted from initial conditions A (dashed lines) and calculated from simulations (colored symbols). The dotted lines are the approximate predictions keeping only the term $\tau_S$. In this example $R_p=2$.}
    \label{fig:mean_duration_A_well}
\end{figure}

The agreement between the predictions of this expression and the simulated data is quite good (dashed lines in Fig. \ref{fig:mean_duration_A_well}). Of the four terms, one [called $\tau_S$ in Eq. (\ref{eq:tauS})] is dominant when $E_w\gg kT$, i.e. in typical situations when the contact is stable, because it is proportional to $e^{2\beta E_w}$ (Fig. S6). Keeping only this term, the agreement with the simulated data is worse (dotted line in Fig. \ref{fig:mean_duration_A_well}), but still useful considering the great simplification it implies. In the limit of $a\ell^{1/2}\gg\epsilon$, it converges as $\sim \ell^{-1}$ to
\begin{equation}
    \overline{\tau}=\frac{R_p^2(\epsilon-R_p)(e^{\beta E_w}-1)}{3D\epsilon [\epsilon^3+R_p^3(e^{\beta E_w}-1)]},
    \label{eq:tauAintApp}
\end{equation}
that can be used to obtain the values of $E_w$ and $R_p$ from experiments where the values of $\overline{\tau}$ associated to at least two different $\epsilon$ are observed.

Note that this approximation fails at $\epsilon\to R_p$, for which a different treatment is necessary (Sect. \ref{sect:erp} below).

\subsection{Initial conditions of type B}

If the residence time is calculated from a single continuous trajectory and if $R_p\ll\epsilon$, the distribution $p_0(R)$ has the same form as that of the ideal chain [Eq. (\ref{eq:p_0_B_approx})]. In fact, in this case the jumps that make the contact accumulate close to $\epsilon$, where the interaction between the ends of the chain does not yet act. 

As a consequence, we can also approximate $\tau(R)$ [Eq. (\ref{eq:tau01})] in the neighbourhood of $\epsilon$ [Eq.(\ref{eq:tau_B_well_approx})] and integrate them, obtaining
\begin{eqnarray}
\overline{\tau}&\approx& \overline{\tau}_0+\frac{a^2\ell\sqrt{\pi\Delta t}}{12\epsilon^2 \sqrt{D}} (e^{\beta E_w}-1)\times\nonumber\\
&&\left( \sqrt{\frac{2\pi}{3}}a\ell^{1/2}e^{\frac{3\epsilon^2}{2a^2\ell}} \text{erf}\left(\frac{3R_p}{2a\ell^{1/2}}\right)-2 R_p e^{\frac{3(\epsilon^2-R_p^2)}{2a^2\ell}}\right),
\label{eq:tauBint}
\end{eqnarray}
where $\overline{\tau}_0$ is the mean residence time of the ideal chain [Eq. (\ref{eq:tauB2})]. Both these two terms behave as $\sim \frac{1}{\ell}$ for large $\ell$, meaning that the asymptotic behaviour of the mean residence time is qualitatively the same of the ideal chain.

\begin{figure}
    \includegraphics[width=\linewidth]{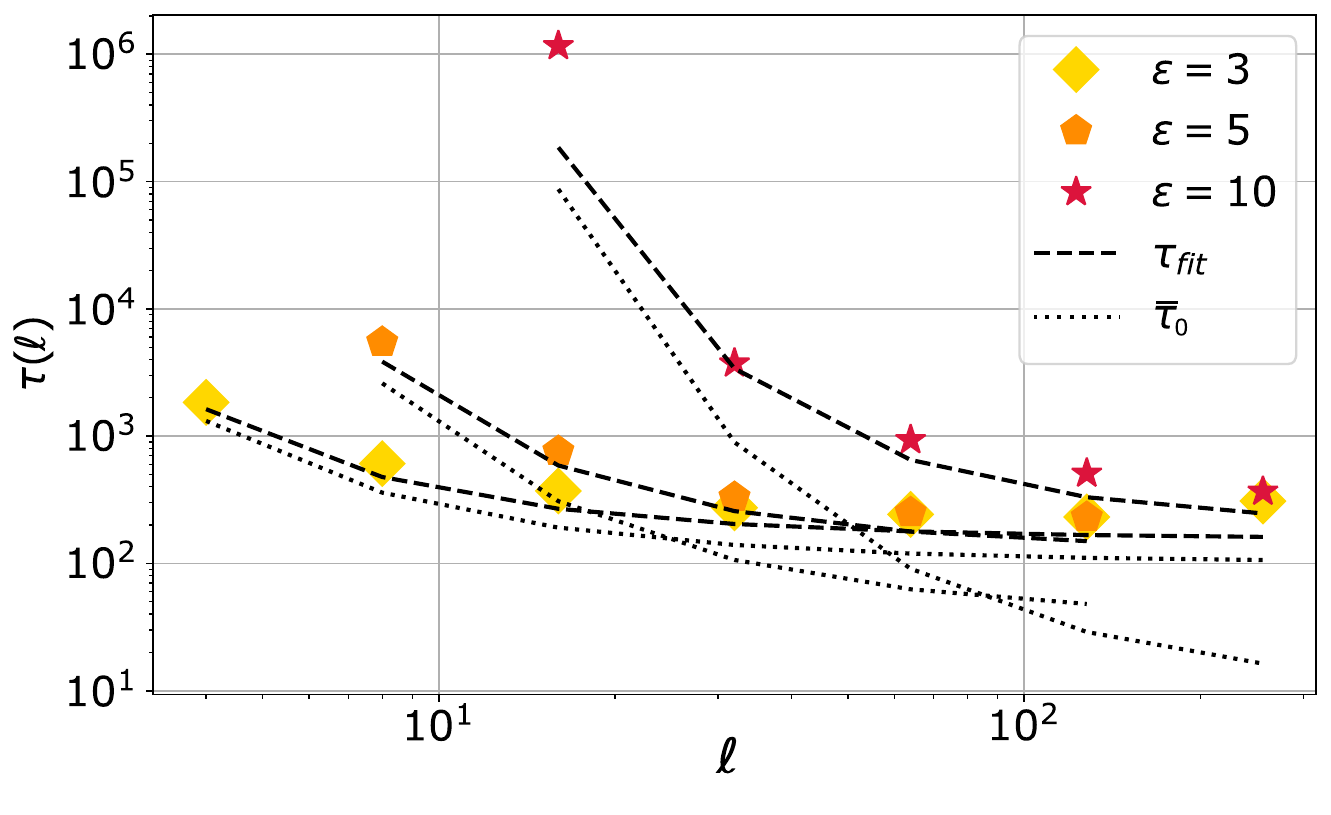}
    \caption{Mean residence time of contacts of the interacting chain as a function of $\ell$ predicted for initial conditions of type B (dashed line, $\Delta t = 2.5$) and calculated from simulations (coloured symbols). The dotted curves are the approximate predictions obtained by keeping only the term related to $\tau_0(R)$.} 
    \label{fig:mean_duration_B_well}
\end{figure}

Equation (\ref{eq:tauBint}) agrees quite well with the simulated residence times (dashed lines in Fig. \ref{fig:mean_duration_B_well}). If $\epsilon$ is small enough, the term $\overline{\tau}_0$ can be neglected (dotted lines in Fig. \ref{fig:mean_duration_B_well}). In this case, the second term of Eq. (\ref{eq:tauBint}) can be used to obtain the values of $E_w$ and $R_p$ from an experimental trajectory, collecting the residence times for two different values of $\epsilon$.

In the case $a\ell^{1/2}\gg\epsilon$, the residence time is even simpler and gives
\begin{equation}
    \overline{\tau} \approx \overline{\tau}_0 + \frac{R_p^3\sqrt{\pi\Delta t}}{\epsilon^2 \sqrt{D}} (e^{\beta E_w}-1),
\end{equation}
where the dependence on $\ell$ in the second term cancels out.

The curves defined by $\overline{\tau}$ at different values of $\epsilon$ (e.g. $\epsilon=3$ and $\epsilon=5$ in Fig. \ref{fig:mean_duration_B_well}) may overlap. In fact, $\overline{\tau}$ as a function of $\epsilon$ [Eq. (\ref{eq:tau_R_potential_2})] is non-monotonic for any given $R$, with a local maximum at $\epsilon_{\text{max}} = R_p+\Delta R$ (Fig. S7), 
where $\Delta R = \sqrt{4 D dt}$.
In other words, for large $\epsilon$ the residence time can be long because the system can sample a larger space, while for small $\epsilon$ it can be long because it is easier to fall into the energy well.

\subsection{Case $\epsilon\approx R_p$} \label{sect:erp}

If we know the width of the energy well, we can study the mean residence time using $\epsilon\approx R_p$. This case is tricky because the potential we are using has a discontinuity in $R_p$. If we set $\epsilon=R_p$, we are predicting the time needed to reach the boundary of the well without crossing it. In fact, under this condition $\overline{\tau}$ would give the same expression $\tau_0(R)$ as the ideal chain, independent of $E_w$. Moreover, the approximation of Eq. (\ref{eq:tauAintApp}) no longer works because it goes to zero in the limit of $\epsilon \to R_p$.

One way to get reasonable results without increasing the complexity of the potential function is to set $\epsilon=R_p+\delta R$, where $\delta R>0$. Assuming that the well is deep enough ($E_w\gg kT$) and that $R_p\ll a\ell^{1/2}$, we obtain [Eq. (\ref{eq:tauAintApp})]
\begin{equation}
    \overline{\tau}=\frac{1}{3D}\frac{R_p^2\,\delta R}{\delta R+R_p}\,e^{\beta E_w},
    \label{eq:arrehnius}
\end{equation}
which has the form of the Arrhenius equation and describes the opening of the contact as a thermally activated event, independent of the length of the polymer. The value of $\delta R$ has a lower bound given by the validity condition of the Fokker-Planck equation, of which Eq. (\ref{eq:kramers}) is the solution, namely $\delta R\gtrsim (Dm/\gamma)^{1/2}$, where $m$ is the mass and $\gamma$ is the coefficient of friction of the monomer; this value corresponds to the distance travelled by the monomer due to its inertia, which is otherwise neglected in our treatment. If the well is deep and thin ($R_p\to 0$), the mean residence time converges to 
\begin{equation}
    \overline{\tau}=\frac{R_p^2\,e^{\beta E_w}}{3D},
\end{equation}
independent of $\delta R$.

An interesting question is what is the value of $\epsilon=R_p$ for which the Arrhenius--like behaviour (\ref{eq:arrehnius}) holds. Simulations with $\epsilon<2$ are computationally demanding because it is difficult to collect statistics. For $R_p=\epsilon=2$ we still observe a weak dependence on $\ell$ (Fig. \ref{fig:mean_duration_A_well}) and is thus an upper bound to the Arrhenius-like behavior. We then suggest that this occurs when $R_p$ is of the order of the microscopic lengthscale of the system $a=1$, defined by the rest distance between consecutive monomers.

For $p_0(R)$ corresponding to initial conditions of type B [Eq. (\ref{eq:p_0_B_approx})], we face a similar problem and we solve it similarly to the case of initial conditions of type A, defining $\epsilon=R_p+\delta R$ and taking the limit of $E_w\gg kT$ [Eq. (\ref{eq:tauBint2})]. For $a\ell^{1/2}\gg R_p$ we find the same result as that of initial conditions A [Eq. (\ref{eq:arrehnius})]. This is not unexpected, because if the well dominates the dynamics and we use the correct definition of its width, the diffusion of the monomers within the well becomes irrelevant.

\section{Discussion and Conclusions}

The residence time of contacts in polymers is an important feature of many biological systems \cite{Copeland2006,Yang2024EnhancerActivation}, although it is difficult to measure in single molecule experiments due to the lack of a precise definition of a contact length based on a molecular understanding of the systems of interest. We modelled the mean residence time $\overline{\tau}$ in two simple cases, that of an ideal chain and that of a chain with attracting ends, studying how it depends on the definition of a contact. For a strongly interacting polymer, $\overline{\tau}$ is that given by Arrhenius equation only if the defined contact length is equal to the width of the energy well that causes the interaction and if the polymer is long enough. Anyway, we obtain some equations that allows one to recover the energy and the width of the well from measurements with arbitrary contact definition.

Even in the simple cases of a non--interacting chain and a polymer with attracting ends, some approximations are necessary to obtain an analytic expression for the mean residence time. Among them, there is the assumption that the motion of the distance $R$ that defines a contact is the slowest dynamical mode of the system, so that one can describe its dynamics as a Markovian diffusion, and the approximation of neglecting the inertia of the system, using the Kramers formalism that derives from overdamped Langevin equations. The agreement between the predictions of the theory and the results of the simulations, which are not affected by these approximations, suggests that the approximations made are valid. Moreover, in all the cases we have analysed, the agreement with the simulation worsen as $\epsilon$ is increased. This a consequence of the fact that in multiple points of the derivation, we assumed $\epsilon\ll R_0$, meaning that the length that defines a contact is much smaller than the equilibrium distance. This is indeed the typical case, in which the entropic force tends to disrupt the contact.

The mean residence time is not an intrinsic property of the system, but depends on the initial conditions of the specific experiment. We tested two initial conditions, where the chain is either in an equilibrium state conditioned to the formation of the contact, or the chain has just formed a contact. The latter, in particular, is the typical case of an experiment in which a long dynamical trajectory is studied by identifying when the contact is formed and is broken. To be noted that in this case, the result depends on the time interval $\Delta t$ between the observations, and thus on the observer. The mean residence times have a different functional form with respect to the parameters of the system for the two types of initial conditions.

The mean residence time has in general a decreasing dependence on the length $\ell$ of the polymer that saturates for large $\ell$. This combines the dependence on $\ell$ of the residence time for any given initial point, caused by molecular diffusion within the definition range of the contact, with that of the distribution of initial points. We find repeatedly a $\ell^{-1}$ dependence for both types of initial conditions we considered. This means that it affects even very long polymers, or pairs of monomers that are very distant along the chain.

Different scaling laws with scaling exponents between $-2$ and $-0.5$ are found in experiments \cite{Shen2001,Kuznetsov2008}. We found an intermediate value for simple homopolymeric models that do not describe the complex network of system dependent interactions that stabilise the experimental system. Therefore, we suggest that the exponents found experimentally are not a general polymer property, but are associated with the specific molecular system. In other words, the exponent $-1$ is that expected in a polymeric contact if no other mechanisms are present. It is the reference exponent that can be used to evaluate from experiments where $\ell$ is varied if other specific mechanisms are at work.

The $\ell$--dependence of the residence time arises when there is a dynamics within the length that defines the contact. On the other hand, the approximation sometimes used in Markov models, which describes the opening of a contact as a length independent Poisson process, is only reasonable under the strict hypothesis that the bound state is unique.

\appendix

\section{Simulation details} \label{sect:details}

All the simulations are performed using the LAMMPS software \cite{Thompson22} for molecular dynamics. The dynamics of an ideal chain is simulated using $N$ beads with fixed mass $m = 10^3$ and the interaction with the background implicit solvent is modelled with the Langevin equations \cite{Schneider1978},
\begin{equation}
    m\ddot{r_i} = f_i -\gamma \dot{r_i} + \eta_i(t),
\end{equation}
where $f_i$ is a component of the inter-particle forces, $\gamma$ is the friction coefficient, $\eta_i(t)$ is a random force satisfying extracted from a Gaussian distribution with zero average and $\overline{\eta_i(t)\eta_j(t')}=2\gamma k_BT\delta(t-t')\delta_{ij}$. The Langevin equation is integrated numerically using a velocity--Verlet algorithm with time step $\delta t=0.25$. Both the  friction coefficient and the mass of the particles is set equal to $10^3$ (in their respective LAMMPS units). The temperature is set such that $k_BT=1$. In this way, the damping time is $\tau_d=m/\gamma=1$ and the diffusion coefficient $D=k_BT/\gamma=10^{-3}$.

Each bead (excluding the head and tail of the chain) is bonded to the two neighbouring beads via a harmonic potential with constant $k=100$ and rest length $a = 1$. Simulations are initiated from random conformations with a Maxwell--Boltzmann distribution of velocities. We performed 20 independent simulations of an ideal chain of length ($N=1000$ beads), studying the dynamics of all pairs of monomers separated by linear length $\ell\leq N$. We checked that this is equivalent to monitoring the distance between pairs of monomers at the ends of polymers of length $\ell$ (cf. Fig. S8 in the Supp. Mat.). Each simulation is equilibrated for $10^7$ steps; after equilibration, the dynamics is simulated for additional $10^5$ steps, saving the conformation every $10$ steps. 

In the case of a polymer with interacting ends we performed a single long simulation of $10^8$ steps for each polymer of length $\ell$ (saving the trajectory every $10$ steps). In this case, we added to the ideal chain an  attractive interaction between the ends of the polymer with the form (Fig. S9 in the Supp. Mat.)
\begin{equation}
\label{eq:sigmoid}
    V(R) = \frac{E_w}{1+e^{-\frac{R-R_p}{b}}}- E_w,
\end{equation}
where $\beta E_w$ sets the depth of the well and $a$ defines the width of the increasing region. The value of the potential depth has been set to $E_w = 2$ in our simulation. This value is low enough to make the opening and closing of the ring frequent while still affecting significantly the typical duration of contacts.
Finally, the value of the width parameter has been set to $b = 0.01$, small enough to make the potential effective only in a small region around $R_p$ but high enough to avoid instabilities with the numerical integration.

We did not try to transform LAMMPS units to real--life ones, since this study is rather general and we do not describe a specific molecular system. Anyway, if we deal with DNA or RNA hairpins, we can interpret the lengths in our simulations in nanometers ($a\approx$ 0.5 nm) and the times in microseconds ($\Delta t\approx 0.25\;\mu$s). In this way, the simulated residence times of the contacts are of the same order as those found in experiments \cite{Shen2001,Kuznetsov2008}.

The distribution of residence times (Fig. 1) depends on $\ell$ for both initial conditions A and B. Typically, it has a power--law  distribution $t^{-\alpha}$ with small exponent ($\alpha\lesssim 1$) at short times and an exponential cutoff, making the mean residence time finite.

\section{Effective model for the ideal chain} 
\label{sect:effective}

The effective force $\mathbf{f}(\vR)$ that acts between pairs of monomers is due to the entropy of the polymer.
The density of the end--to--end difference vectors $\vR$ in an ideal chain, arising from the central limit theorem, is
\begin{equation}
    \Omega(\vR)=\left(\frac{2\pi\ell a^2}{3}\right)^{-3/2}\exp\left[-\frac{3|\vR|^2}{2\ell a^2}, \right],
    \label{eq:eq_ideal}
\end{equation}
which produces an effective energy
\begin{equation}
    \tilde{U}(\vR)=-kT\log\Omega(\vR)=kT\frac{3|\vR|^2}{2\ell a^2}+\text{const}
    \label{eq:effU}
\end{equation}
and thus an effective force $\mathbf{\tilde{f}}=-{\bf{\nabla}}\tilde{U}$.

The Fokker--Planck equation that controls the probability $\tilde{p}(\vR,t)$ as a function of the  vector $\vR$ is
\begin{equation}
    \frac{\partial \tilde{p}(\vR,t)}{\partial t}=-\frac{D}{kT}\nabla\left(\tilde{\mathbf{f}} \;\tilde{p}(\vR,t) \right) +D\nabla^2  \tilde{p}(\vR,t).
\end{equation}
In spherical coordinates $(R,\theta,\phi)$, the Fokker--Planck equation can be written as
\begin{equation}
    \frac{\partial \tilde{p}(R,\theta,\phi,t)}{\partial t}=-\frac{D}{kT\,R^2}\frac{\partial}{\partial R} [R^2 \tilde{f}(R,\theta,\phi) \tilde{p}(R,\theta,\phi,t)]+D\frac{1}{R^2}\frac{\partial}{\partial R} [R^2 \frac{\partial}{\partial R} \tilde{p}(R,\theta,\phi,t)]+\dots,
\end{equation}
where the dots indicate the derivatives with respect to $\theta$ and $\phi$ and $\tilde{f}$ is the radial component of the force.
Since the force associated with Eq. (\ref{eq:effU}) depends only on the radial coordinate, one can study the probability as a function of the scalar $R=|\vR|$ only,
\begin{equation}
    p(R,t)=4\pi R^2 \tilde{P}(|\vR|,t),
    \label{eq:pr_1d}
\end{equation}
thus
\begin{equation}
    \frac{1}{4\pi R^2}\frac{\partial p(R,t)}{\partial t}=-\frac{D}{kT\,R^2}\frac{\partial}{\partial R} \left[R^2 \tilde{f}(R) \frac{p(R,t)}{4\pi R^2} \right]+D\frac{1}{R^2}\frac{\partial}{\partial R} \left[R^2 \frac{\partial}{\partial R} \frac{p(R,t)}{4\pi R^2}\right].
\end{equation}
Multiplying by $4\pi R^2$ and performing the inner derivative in the diffusive term,
\begin{equation}
    \frac{\partial p(R,t)}{\partial t}=-\frac{D}{kT}\frac{\partial}{\partial R} [\tilde{f}(R) p(R,t) ]+D\frac{\partial}{\partial R} \left[ -\frac{2}{R}p(R,t)+\frac{\partial p(R,t)}{\partial R} \right],
\end{equation}
and eventually the  1--dimensional equation
\begin{equation}
    \frac{\partial p(R,t)}{\partial t}=-\frac{D}{kT}\frac{\partial}{\partial R} \left[\left( \tilde{f}(R)+\frac{2kT}{R} \right)p(R,t)\right]+D\frac{\partial^2}{\partial R^2} p(R,t),
\end{equation}
which contains the spurious drift $2kT/R$, which can be regarded as derivative of the effective potential $-kT\log R^2$. This equation can be rewritten as
\begin{equation}
    \frac{\partial p(R,t)}{\partial t}=-\frac{D}{kT}\frac{\partial}{\partial R} \left[ f(R) p(R,t)\right]+D\frac{\partial^2}{\partial R^2} p(R,t),
\end{equation}
with $f(R)=-dU/dR$ and
\begin{equation}
\label{eq:potential}
    U(R)=kT\left[\frac{3R^2}{2\ell a^2}-\log R^2\right].
\end{equation}


\section{Residence time for the ideal chain}
\label{sect:residence_supp}

We start from the solution of Kramers equation for a 1--dimensional system in the radial coordinate 
\begin{equation}
    \tau(R)=\frac{1}{D}\int_R^\epsilon dR'\; e^{\beta U(R')}\int_0^{R'}dR''\; e^{-\beta U(R'')},
\end{equation}
that we solve using the potential of Eq. (\ref{eq:potential}),
\begin{equation}
\label{eq:suppTauR}
\begin{split}
     \tau(R)& =\frac{1}{D}\frac{R_0^2}{4}\int_R^\epsilon dR' \left [ -2R'\exp\left(-\frac{R'^2}{R_0^2}\right)+\sqrt{\pi}R_0\text{erf}\left( \frac{R'}{R_0}\right) \right]\frac{\exp\left(\frac{R'^2}{R_0^2}\right)}{R'^2} \\
     & = \frac{1}{6D}\left[\epsilon^2\; _2F_2\left( \frac{\epsilon^2}{R_0^2} \right)-R^2\; _2F_2\left( \frac{R^2}{R_0^2} \right)  \right],
\end{split}
\end{equation}
where we defined $R_0 = \sqrt{\frac{2\ell a^2}{3}}$.

For initial conditions of type A, we have to integrate $\tau(R)$ with the distribution of Eq. (\ref{eq:pr_1d}),
\begin{equation}
     \overline{\tau}=\frac{4\pi}{6DA(\epsilon, \ell)}\left(\frac{1}{\pi R_0^2} \right)^{3/2}\int_0^\epsilon dR\,\left[\epsilon^2 \,_2F_2\left(\frac{\epsilon^2}{R_0^2}\right)-R^2\,_2F_2\left(\frac{R^2}{R_0^2}\right)\right] R^2 e^{-\frac{R^2}{R_0^2}}.
\end{equation}
where we defined 
\begin{equation}
\begin{split}
    A(\epsilon, \ell) & = 4\pi\left(\frac{1}{\pi R_0^2} \right)^{3/2}\int_0^\epsilon dR\, R^2\exp\left[-\frac{R^2}{R_0^2}\right] = \\ & =  \frac{1}{\sqrt{\pi}R_0} \left(-2\epsilon \exp\left[-\frac{\epsilon^2}{R_0^2}\right]+\sqrt{\pi}R_0 \text{erf}\left[ \frac{\epsilon}{R_0} \right].\right)  
\end{split}    
\end{equation}
For small $\epsilon/R_0$, the approximation $_2F_2(R^2/R_0^2)\approx 1+\frac{R^2}{5R_0^2}$ is rather good down to $R_0\approx\epsilon$ (cf. Fig. S10), and one can obtain

\begin{equation}
\label{eq:tau_th_A}
    \overline{\tau}=\frac{1}{12D\sqrt{\pi}R_0 A(\epsilon, \ell)}\left( \epsilon(9R_0^2+2\epsilon^2)\exp\left( -\frac{\epsilon^2}{R_0^2}\right)-\frac{\sqrt{\pi}(45R_0^4-20R_0^2\epsilon^2-4\epsilon^4)}{10 R_0}\text{erf}\left[\frac{\epsilon}{R_0}\right] \right),
\end{equation}
which for $\epsilon\ll R_0$, to the fifth order in $\epsilon/R_0$, is
\begin{equation}
     \overline{\tau}\approx \frac{\epsilon^2}{15D}\left( 1+\frac{16}{35}\frac{\epsilon^2}{R_0^2} \right),
\end{equation}
while for $\epsilon\approx R_0$ is
\begin{equation}
    \overline{\tau}\approx \frac{21h-110}{120(h-2)}\frac{\epsilon^2}{D} - \frac{49h^2-222h+112}{60(h-2)^2}\frac{(R_0-\epsilon)\epsilon}{D}\approx \frac{0.10\epsilon^2}{D}-0.07\frac{(R_0-\epsilon)\epsilon}{D},
\end{equation}
where $h\equiv \sqrt{\pi}\cdot e\cdot\text{erf}(1)$. Rewriting the expression as a function of $\ell$ we get 
\begin{equation}
    \overline{\tau}\approx \frac{0.17\epsilon^2}{D}-0.06\frac{\epsilon a \sqrt{\ell}}{D}.
    \label{eq:tauA_apf}
\end{equation}

For initial conditions of type B, we have to calculate the distribution of initial distances $p_0(R)$. We employed the Chapman--Kolmogorov equation (8) 
\begin{equation}
    p_0(R)=\int_0^\infty dR'\, p_{\Delta t}(R,t+\Delta t|R',t)p(R',t|R'>\epsilon,t),
    \label{eq:supp-chapman}
\end{equation}
using the conditional probability
\begin{equation}
    p(R',t|R'>\epsilon,t)=\frac{4}{\sqrt{\pi}R_0^3}
    \frac{R'^2\exp\left[-\frac{R'^2}{R_0^2} \right]\theta(R'-\epsilon)}
    {\int_\epsilon^\infty dR\, R^2\exp\left[-\frac{R^2}{R_0^2}\right]} =
    \frac{4}{\sqrt{\pi}R_0^3}\frac{R'^2\exp\left[-\frac{R'^2}{R_0^2} \right]\theta(R'-\epsilon) }{1-A(\epsilon, \ell)}
    \label{eq:cond1}
\end{equation}
obtained from the stationary distribution of the distances in the ideal chain, properly normalised. Here we defined $\theta$ a step function that forces the conditional probability to zero for distances at which the contact is already formed.

The maximum of the conditional probability is at
\begin{equation}
    R'=
    \begin{cases}
    \epsilon &\text{if $\epsilon<R_0$ } \\
    R_0 &\text{if $\epsilon\geq R_0$ }.
    \end{cases}
\end{equation}

Furthermore, we used the Onsager--Machlup conditional probability
\begin{equation} 
    p_{\Delta t}(R,t+\Delta t|R',t)=\frac{1}{(4\pi D\Delta t)^{1/2}}\exp\left[-\frac{(R'-R-f(R')D\Delta t/kT)^2}{4D\Delta t}\right]
    \label{eq:onsager}
\end{equation}
that holds if $\Delta t$ is small with respect to the motion timescale of the system, $\Delta t\ll \min[kT\epsilon/fD),\epsilon^2/D]$. 

The Chapman--Kolmogorov integral cannot be calculated simply using Eq. (\ref{eq:onsager}). However, for large $\ell$ and small $\epsilon$,
the maximum of Eq. (\ref{eq:cond1}) is the very point at which $f(R')=0$, so we can neglect the last term in the exponential of Eq. (\ref{eq:onsager}). Moreover, Eq. (\ref{eq:cond1}) can be rewritten as
\begin{equation}
    p(R',t|R'<\epsilon,t)=\frac{4}{\sqrt{\pi}R_0}
     \frac{\exp\left[-\frac{R^2}{R_0^2}+2\log\frac{R'}{R_0}\right]\theta(R'-\epsilon)}{1-A(\epsilon, \ell)}.
    \label{eq:cond2}
\end{equation}
Now, one can write $\log\frac{R'}{R_0}=\log\left[1+\frac{R'-R_0}{R_0} \right]$, that can be approximated with $\frac{R'-R_0}{R_0}$ close to the maximum of the conditional probability. Thus,
\begin{equation}
    p(R',t|R'<\epsilon,t)\approx\frac{4}{e\sqrt{\pi}R_0}\frac{\exp\left[-\left (\frac{R'-R_0}{R_0} \right)^2 \right]\theta(R'-\epsilon)}{1-A(\epsilon, \ell)}
    \label{eq:cond3}
\end{equation}
and
\begin{equation}
    p_0(R)=
    \frac{2}{e\pi R_0(D\Delta t)^{1/2}(1-A(\epsilon, \ell))}
    \int_\epsilon^\infty dR'\, 
    \exp\left[
    -\left(\frac{R'-R_0}{R_0}\right)^2 
    -\frac{(R'-R)^2}{4D\Delta t}
    \right]
\end{equation}
This can be integrated exactly because it has a Gaussian integrand. However, one can give an estimate of the integrand in the limit of small $\Delta t$ because in this case the only jumps that can fall within $R<\epsilon$ according to Eq. (\ref{eq:onsager}) are those starting close to $\epsilon$. Then, the latter term in the exponent is much larger than the former (for $R<R'$) and, normalizing accordingly,
\begin{equation}
    p_0(R)\approx \frac{\sqrt{\pi}}{\sqrt{4D\Delta t}} \left[1-\text{erf}\left(\frac{\epsilon-R}{(4D\Delta t)^{1/2}} \right) \right],
    \label{eq:suppp0r}
\end{equation}
which is independent on $\ell$. This appears as a good approximation as compared to simulations for $\ell\gg 1$ and $\epsilon\sim 1$ (Fig. S11).

The mean residence time can be calculated integrating 
\begin{equation}
    \overline{\tau}=\int_0^\epsilon dR\, \tau(R) p_0(R),
\end{equation}
together with Eqs. (\ref{eq:suppTauR}) and (\ref{eq:suppp0r}). This gives
\begin{equation}
    \overline{\tau}= \frac{\sqrt{\pi}}{12\sqrt{D^{3}\Delta t}}\int_0^\epsilon dR\, 
   \left[\epsilon^2\; _2F_2\left( \frac{\epsilon^2}{R_0^2} \right)-R^2\; _2F_2\left( \frac{R^2}{R_0^2} \right)  \right]
  \left[1-\text{erf}\left(\frac{\epsilon-R}{(4D\Delta t)^{1/2}} \right) \right].
\end{equation}
Both integration terms are peaked around $\epsilon$. The former can be approximated around $\epsilon$, defining
\begin{equation}
    g(R)=R^2\cdot\,_2F_2\left(\frac{R^2}{R_0^2}\right),
\end{equation}
under the assumption that $\Delta t$ is so small that only small jumps into the region $R<\epsilon$ can occur from outside, as
\begin{equation}
    \left.\frac{dg}{dR}\right|_{R=\epsilon}\cdot(\epsilon-R)=
    \frac{3R_0^2}{2\epsilon^2}\left[ \sqrt{\pi}R_0e^{\frac{\epsilon^2}{R_0^2}} \text{erf}\left(\frac{\epsilon}{R_0}\right)-2\epsilon\right](\epsilon-R).
\end{equation}
The integral is then, for small $\Delta t$ (i.e., when $\sqrt{D\Delta t}\ll\epsilon$),
\begin{equation}
\label{eq:supp_tau_final}
    \overline{\tau}\approx \frac{R_0^2\sqrt{\pi\Delta t}}{8\epsilon^2 \sqrt{D}}
    \left[ 
    \sqrt{\pi}R_0e^{\frac{\epsilon^2}{R_0^2}} \text{erf}\left(\frac{\epsilon}{R_0}\right)-2\epsilon
    \right],
\end{equation}
which, in the limit of large $\ell$, scales as
\begin{equation}
    \overline{\tau}\approx \frac{\sqrt{\pi \Delta t}\epsilon}{6\sqrt{D}}\left(1 + \frac{2}{5}\frac{\epsilon^2}{R_0^2}\right).
\end{equation}

\section{Ideal chain with attracting ends}

Let's introduce an attractive interaction between the ends of the polymer, modelled as a square well
\begin{equation}
    V(R) = \begin{cases}
        -E_w & \text{if } R < R_p \\
        0 & \text{if } R \geq R_p,
    \end{cases}
\end{equation}
so that the total potential is now
\begin{equation}
U(R)+V(R) = \begin{cases} kT\left[\frac{3R^2}{2\ell a^2}-\log R^2 -\beta E_w\right]  & \text{if } R < R_p \\ 
kT\left[\frac{3R^2}{2\ell a^2}-\log R^2 \right] & \text{if } R \geq R_p, \\
\end{cases}
\end{equation}
where $\beta\equiv 1/k_BT$. For $R_p>\epsilon$ the potential has no effect on the residence time because the potential is constant within the boundaries of the integrals of Eq. (\ref{eq:kramers}) and the results already obtained for the ideal chain will still hold. In the opposite case, it is useful to split the integrals in the two cases $R<R_p$ and $R\geq R_p$. In the former case we have
\begin{equation}
\begin{split}
    \tau(R) & = \frac{1}{D}\bigg[\int_R^{R_p}dR' e^{-\beta E_w}P^{-1}(R')\int_0^{R'}dR''e^{\beta E_w}P(R'') + \\ & + \int_{R_p}^{\epsilon}dR'P^{-1}(R')\left (\int_0^{R_p}dR''e^{\beta E_w}P(R'') + \int_{R_p}^{R'}dR''P(R'') \right)\bigg] =\\ 
    & = \frac{1}{D}\left[\int_R^{R_p}dR' P^{-1}(R')\int_0^{R'}dR''P(R'') + \right.\\
    &+\left.\int_{R_p}^{\epsilon}dR'P^{-1}(R')\left (\int_0^{R_p}dR''e^{\beta E_w}P(R'') +\int_{0}^{R'}dR''P(R'') - \int_{0}^{R_p}dR''P(R'')\right)\right] =\\
    & = \tau_0(R) +\frac{e^{\beta E_w}-1}{D}\int_{R_p}^{\epsilon}dR'P^{-1}(R')\int_0^{R_p}dR''P(R'')
\end{split}    
\end{equation}
while when $R\geq R_p$ we obtain
\begin{equation}
\begin{split}
 \tau(R) & = \frac{1}{D}\int_R^{\epsilon}dR'P^{-1}(R')\left (\int_0^{R_p}dR''P(R'')e^{\beta E_w} + \int_{R_p}^{R'}dR''P(R'') \right) =\\ 
 & = \frac{1}{D}\int_R^{\epsilon}dR'P^{-1}(R')\left( \int_0^{R_p}dR''P(R'')e^{\beta E_w} + \int_{0}^{R'}dR''P(R'') - \int_{0}^{R_p}dR''P(R'') \right) =\\ 
 & =  \tau_0(R) +\frac{e^{\beta E_w}-1}{D}\int_{R}^{\epsilon}dR'P^{-1}(R')\int_0^{R_p}dR''P(R''),
\end{split}
\end{equation}
where we worked on the integration boundaries to isolate the residence time $\tau_0$ of the ideal chain, in absence of the attracting potential. Summing up,
the residence time of the contact as a function of $R$ can be written as
\begin{equation}
     \tau(R) = \begin{cases}
        \tau_0(R) +\frac{e^{\beta E_w}-1}{D}\int_{R_p}^{\epsilon}dR'P^{-1}(R')\int_0^{R_p}dR''P(R'') & \text{if } R < R_p \\ 
         \tau_0(R)+\frac{e^{\beta E_w}-1}{D}\int_{R}^{\epsilon}dR' P^{-1}(R')\int_0^{R_p}dR''P(R'')  & \text{if } R \geq R_p.
     \end{cases}
\end{equation}
When  $R<R_p$, the second term is independent on $R$. Therefore, in this case the average residence time is that of an ideal chain plus a constant that depends only on $\epsilon$ and $\ell$. 
The integrals can be evaluated exactly, giving
\begin{equation}
\begin{split}
\label{eq:tau_R_potential_1}
     \tau(R) & = \tau_0(R) + \frac{e^{\beta E_w}-1}{4D}R_0^2\left( \sqrt{\pi}\text{erf}\left(\frac{R_p}{R_0}\right)-2\frac{R_p}{R_0}e^{-\frac{R_p^2}{R_0^2}}\right)\bigg[ \sqrt{\pi}\erfi\left(\frac{\epsilon}{R_0}\right)- \\ & + \sqrt{\pi}\erfi\left(\frac{R_p}{R_0}\right)-\frac{R_0}{\epsilon}e^{\frac{\epsilon^2}{R_0^2}} + \frac{R_0}{R_p}e^{\frac{R_p^2}{R_0^2}}\bigg]
\end{split}     
\end{equation}
for $R< R_p$ and 
\begin{equation}
\begin{split}
\label{eq:tau_R_potential_2}
     \tau(R) & = \tau_0(R) + \frac{e^{\beta E_w}-1}{4D}R_0^2\left( \sqrt{\pi}\text{erf}\left(\frac{R_p}{R_0}\right)-2\frac{R_p}{R_0}e^{-\frac{R_p^2}{R_0^2}}\right)\bigg[ \sqrt{\pi}\erfi\left(\frac{\epsilon}{R_0}\right)- \\ & + \sqrt{\pi}\erfi\left(\frac{R}{R_0}\right)-\frac{R_0}{\epsilon}e^{\frac{\epsilon^2}{R_0^2}} + \frac{R_0}{R}e^{\frac{R^2}{R_0^2}}\bigg]
\end{split}     
\end{equation}
when $R \geq R_p$. The residence time of the contact can thus be written in the form
\begin{equation}
    \tau(R) = \tau_0(R) + \tau_1(R),
    \label{eq:tau01}
\end{equation}
where we defined
\begin{equation}
\begin{split}
    \tau_1(R) = 
        \frac{e^{\delta}-1}{4D}R_0^2\left( \sqrt{\pi}\text{erf}\left(\frac{R_p}{R_0}\right)-2\frac{R_p}{R_0}e^{-\frac{R_p^2}{R_0^2}}\right)\!\bigg[ \sqrt{\pi} \erfi\left(\frac{\epsilon}{R_0}\right)- \\ + \sqrt{\pi}\erfi\left(\frac{\mathcal{R}}{R_0}\right)-\frac{R_0}{\epsilon}e^{\frac{\epsilon^2}{R_0^2}} + \frac{R_0}{\mathcal{R}}e^{\frac{\mathcal{R}^2}{R_0^2}}\bigg] 
    \label{eq:tau1Rcases}
\end{split}    
\end{equation}
with 
$\mathcal{R}= \text{max}[R, R_p]$.
The residence time $\tau_1(R)$ in the former case, independent of $R$, will be simply written as $\tau_1$.  Examples of these functions are shown in Fig. S12 for different values of the genomic distance $\ell$ and compared with the corresponding curves for the ideal chain. 

It can be instructive to evaluate the residence time in the limit $R_0\to\infty$, which gives
\begin{equation}
\label{eq:tau_R_potential_approx}
      \tau(R) = \begin{cases} 
        \tau_0(R) +\frac{e^{\beta E_w}-1}{2D}R_p^3\left( \frac{1}{R_p}-\frac{1}{\epsilon}\right) & \text{if } R < R_p \\ 
         \tau_0(R)+\frac{e^{\beta E_w}-1}{2D}R_p^3\left( \frac{1}{R}-\frac{1}{\epsilon}\right)  & \text{if } R \geq R_p.
     \end{cases}
\end{equation}

Finally, we have to average $\tau(R)$ over the probability distribution of the initial distances, obtaining two different results for initial conditions A and B. 

For initial conditions of type A, we use as $p_0(R)$ the stationary probability distribution of the polymer, limited to the region $R\leq\epsilon$. The distribution $p_0(R)$ becomes
\begin{equation}
    p_0(R) = \begin{cases} \frac{e^{\beta E_w}R^2}{Z} e^{-\frac{R^2}{R_0^2}} & \text{if } R < R_p \\
    \frac{R^2}{Z} e^{-\frac{R^2}{R_0^2}} & \text{if } R \geq R_p,
    \end{cases}
\end{equation}
where the normalization constant is
\begin{equation}
    Z = \frac{R_0^2\left[ (e^{\beta E_w}-1)(\sqrt{\pi}R_0\text{erf}\left( \frac{R_p}{R_0}\right)-2R_p e^{-\frac{R_p^2}{R_0^2}} ) + \sqrt{\pi}R_0\text{erf}\left( \frac{\epsilon}{R_0}\right)-2\epsilon e^{-\frac{\epsilon^2}{R_0^2}} \right]}{4}.
    \label{eq:zAint}
\end{equation}
Examples of this distribution are displayed in Fig. S13, for two different values of $\ell$.

The integral that gives the mean residence time is composed of four terms,
\begin{equation}
\begin{split}
    \overline{\tau} & = \frac{1}{Z}e^{\beta E_w}\int_0^{R_p}dR  R ^2 e^{-\frac{R^2}{R_0^2}}\tau(R)+\frac{1}{Z}\int_{R_p}^{\epsilon}dR  R ^2 e^{-\frac{R^2}{R_0^2}}\tau(R)= \\ & = \frac{1}{Z}e^{\beta E_w}\int_0^{R_p}dR  R ^2 e^{-\frac{R^2}{R_0^2}}\left(\tau_0(R) + \tau_1\right) + \frac{1}{Z}\int_{R_p}^{\epsilon}dR  R ^2 e^{-\frac{R^2}{R_0^2}}\left(\tau_0(R) + \tau_1(R)\right). 
\end{split}    
\end{equation}
\makebox[\linewidth][s]{The first term is computed using the approximated expression} $\tau_0(R) \approx \frac{1}{6D}\left( \epsilon^2 - R^2 + \frac{1}{5R_0^2}(\epsilon^4 - R^4)\right)$ that holds for $R_0\gg\epsilon$, the same that we used for the ideal chain, obtaining
\begin{equation}
\begin{split}
    \tau_R & =\frac{1}{Z}e^{\beta E_w}\int_0^{R_p}dR  R ^2 e^{-\frac{R^2}{R_0^2}}\tau_0(R) = \frac{1}{Z}\frac{e^{\beta E_w}}{480D}\bigg(2R_p e^{-\frac{R_p^2}{R_0^2}}\big(45R_0^2+ \\ & + 10R_0^2(3R_p^2-2\epsilon^2)+4(R_p^4-\epsilon^4)\big)+ \sqrt{\pi}R_0(45R_0^4 -20R_0^2\epsilon^2-4\epsilon^4)\text{erf}\left(\frac{R_p}{R_0}\right)\bigg).
\end{split}    
\end{equation}    
In the second term, $\tau_1$ does not depend on $R$ and it can be moved out of the integral, giving
\begin{multline}
        \tau_S=\frac{1}{Z}e^{\beta E_w}\tau_1\int_0^{R_p}dR  R ^2 e^{-\frac{R^2}{R_0^2}} = \frac{\frac{1}{Z} e^{\beta E_w}(e^{\beta E_w}-1)R_0^3}{16D}\left(\sqrt{\pi}R_0\text{erf}\left(\frac{R_p}{R_0}\right) -2R_p e^{-\frac{R_p^2}{R_0^2}}\right)^2\cdot \\
        \cdot\left( \sqrt{\pi}\erfi\left(\frac{\epsilon}{R_0}\right)-\sqrt{\pi}\erfi\left(\frac{R_p}{R_0}\right) -\frac{R_0}{\epsilon}e^{\frac{\epsilon^2}{R_0^2}}+\frac{R_0}{R_p}e^{\frac{R_p^2}{R_0^2}} \right).
        \label{eq:tauS}
\end{multline}
The third term is similar to the first, with just different integration boundaries, and gives
\begin{equation}
\begin{split}
        \tau_T &=\frac{1}{Z}\int_{R_p}^{\epsilon}dR  R ^2 e^{-\frac{R^2}{R_0^2}}\tau_0(R) = \\ & = \frac{\frac{1}{Z}}{480D}\bigg(10\epsilon R_0^2e^{-\frac{\epsilon^2}{R_0^2}}(9R_0^2+2\epsilon^2)-2e^{-\frac{R_p^2}{R_0^2}}R_p\big( 45R_0^4 + 10R_0^2(3R_p^2-2\epsilon^2)+ \\ & + 4(R_p^4-\epsilon^4)\big) +\sqrt{\pi}R_0(45R_0^4-20R_0^2\epsilon^2-4\epsilon^4)\left(\text{erf}\left(\frac{R_p}{R_0}\right) -\text{erf}\left(\frac{\epsilon}{R_0}\right)  \right)
        \bigg).
\end{split}        
\end{equation}
To calculate the fourth term, we approximate $\tau_1(R)\approx\frac{e^{\beta E_w}-1}{3D}R_p^3\left(\frac{1}{R}-\frac{1}{\epsilon}\right)$ for $R_0\gg\epsilon$,
 and we  obtain
\begin{equation}
\begin{split}
    \tau_U & = A\int_{R_p}^{\epsilon}\tau_1(R)R^2e^{-\frac{R^2}{R_0^2}}dR = \\ & = \frac{1}{Z}\frac{(e^{\beta E_w}-1)R_p^2R_0^2}{12D\epsilon}\left(2(\epsilon-R_p)e^{-\frac{R_p^2}{R_0^2}}-\sqrt{\pi}R_0\left( \text{erf}\left( \frac{\epsilon}{R_0}\right)-\text{erf}\left( \frac{R_p}{R_0}\right) \right) \right).
    \label{eq:tauU}
    \end{split}
\end{equation}
Putting together all these terms, we obtain an analytical expression 
\begin{equation}
    \overline{\tau}=t_1+t_2+t_3+t_4
    \label{eq:taumA4}
\end{equation}
that we test on the results of simulations in Fig. \ref{fig:mean_duration_A_well}.

For $R_p<\epsilon$ and $E_w\gg k_BT$, the dominant term is always $\tau_S$, corresponding to the integration of $\tau_1$ (which accounts for the contribution of the energy well and is constant for $R<R_p$) between $0$ and $R_p$. However, when $R_p=\epsilon$, the last term between brackets in $\tau_s$ vanishes, and the dominant term becomes $\tau_R$. This is the integral of the contribution of $\tau_0$, that describes the motion as if it were in absence of an energy well. In fact, for $\epsilon\to R_p$ $\tau_R$ is independent on $E_w$. This is not what we want. The residence time we are looking for is that to exit from the energy well, so to reach $\epsilon=R_p+\delta R$. For finite $\delta R$, assuming first that $E_w\gg kT$ we obtain 
\begin{equation}
\begin{split}
    \overline{\tau} & =\frac{e^{\delta} \left(\sqrt{\pi } R_0 \text{erf}\left(\frac{R_p}{R_0}\right)-2 R_p e^{-\frac{R_p^2}{R_0^2}}\right)^2}{4 D \left(\sqrt{\pi }\cdot
   \text{erf}\left(\frac{R_p}{R_0}\right)-\frac{2 R_p}{R_0} e^{\frac{\delta R (\delta R+2 R_p)}{R_0^2}-\frac{(\delta R+R_p)^2}{R_0^2}}\right)} \Biggl[\sqrt{\pi }
   \bigg(\text{erfi}\left(\frac{\delta R+R_p}{R_0}\right) - \\ & + \text{erfi}\left(\frac{R_p}{R_0}\right)\bigg)+ R_0
   \left(\frac{e^{\frac{R_p^2}{R_0^2}}}{R_p}-\frac{e^{\frac{(\delta R+R_p)^2}{R_0^2}}}{\delta R+R_p}\right)\Biggl],
\end{split}   
\end{equation}
that for large $R_0$ is particularly simple,
\begin{equation}
    \overline{\tau}=\frac{R_p^2\,\delta R\,e^{\beta E_w}}{3D\delta R+3DR_p},
    \label{eq:tauAintS}
\end{equation}
and if the energy well is narrow ($R_p\ll\delta R$) gives
\begin{equation}
    \overline{\tau}=\frac{R_p^2\,e^{\beta E_w}}{3D},
\end{equation}
 which has the form of the Arrhenius equation.
 

In the case of initial conditions of type B, we make again use of the $p_0(R)$ of Eq. (\ref{eq:p_0_B_approx}). In fact, since $R_p<\epsilon$, the conditional probability $p(R',t|R'>\epsilon,t)$ of Eq. (\ref{eq:cond1}) does not change with respect to the ideal chain. Moreover, if $R_p\ll\epsilon$, also the jump probability is that of the ideal chain in the region where most of the jumps land, so we can still use the form of $p_0(R)$ found in Eq. (\ref{eq:p_0_B_approx}), fitting its numerical parameters from the simulations, as done for the ideal chain (Fig. S10).

We can now average the residence time of Eqs. (\ref{eq:tau_R_potential_1}) and (\ref{eq:tau_R_potential_2}) over the distribution $p_0(R)$. Since $p_0(R)$ is peaked around $R=\epsilon$, we can use the expression of $\tau(R)$ for $R>R_p$ and we expand it around $R=\epsilon$, obtaining
\begin{equation}
\label{eq:tau_B_well_approx}
    \tau(R) \approx \tau_0(R) + \frac{e^{\beta E_w}-1}{4D}\frac{R_0^2}{\epsilon^2} \left(\sqrt{\pi}R_0e^{\frac{\epsilon^2}{R_0^2}}\text{erf}\left(\frac{R_p}{R_0}\right)-2R_p e^{\frac{\epsilon^2-R_p^2}{R_0^2}}\right).
\end{equation}
This expression can be integrated and we get
\begin{eqnarray}
\overline{\tau}&\approx& \frac{R_0^2\sqrt{\pi\Delta t}}{8\epsilon^2 \sqrt{D}}
    \left[ 
    \sqrt{\pi}R_0e^{\frac{\epsilon^2}{R_0^2}} \text{erf}\left(\frac{\epsilon}{R_0}\right)-2\epsilon + (e^{\beta E_w}-1)\left( \sqrt{\pi}R_0e^{\frac{\epsilon^2}{R_0^2}} \text{erf}\left(\frac{R_p}{R_0}\right)-2 R_p e^{\frac{\epsilon^2-R_p^2}{R_0^2}} \right)\right]=\nonumber\\
    &=&\overline{\tau}_0+\frac{R_0^2\sqrt{\pi\Delta t}}{8\epsilon^2 \sqrt{D}} (e^{\beta E_w}-1)\left( \sqrt{\pi}R_0e^{\frac{\epsilon^2}{R_0^2}} \text{erf}\left(\frac{R_p}{R_0}\right)-2 R_p e^{\frac{\epsilon^2-R_p^2}{R_0^2}}\right),
\end{eqnarray}
where $\overline{\tau}_0$ is the average residence time of the contact for the ideal chain, Eq. (\ref{eq:supp_tau_final}).

In the case $R_0\gg\epsilon$, the residence time is even simpler and gives
\begin{equation}
    \overline{\tau} \approx \overline{\tau}_0 + \frac{4R_p^3\sqrt{\pi\Delta t}}{24\epsilon^2 \sqrt{D}} (e^{\beta E_w}-1),
\end{equation}
where the dependence on $\ell$ in the second term cancels out.

This prediction fails when $\epsilon \approx R_p$; in fact, in this case the starting distance $R$ falls inside the well and we must perform the average over $p_0(R)$ using the definition of $\tau(R)$ for $R<R_p$ (Eq. \ref{eq:tau1Rcases}). In this case, $\tau(R)$ is just $\tau_0(R)$ plus a term that does not depend on R, meaning that no further integration over the distribution of the initial distances is needed and the average contact duration can be written as
\begin{equation}
\begin{split}
    \overline{\tau} & = \overline{\tau}_0+\tau_1 \\
    &  = \overline{\tau_0}+\frac{e^{\beta E_w}-1}{4D}R_0^2\left( \sqrt{\pi}\text{erf}\left(\frac{R_p}{R_0}\right)-2\frac{R_p}{R_0}e^{-\frac{R_p^2}{R_0^2}}\right)\bigg[ \sqrt{\pi}\erfi\left(\frac{\epsilon}{R_0}\right)- \\ & +\sqrt{\pi}\erfi\left(\frac{R_p}{R_0}\right)-\frac{R_0}{\epsilon}e^{\frac{\epsilon^2}{R_0^2}} + \frac{R_0}{R_p}e^{\frac{R_p^2}{R_0^2}}\bigg].
\end{split}
\label{eq:tauBint2}
\end{equation}

In the limit $E_w\gg k_BT$ and for large $R_0$ we obtain
\begin{equation}
\overline{\tau}=\frac{R_p^2\,\delta R\,e^{\beta E_w}}{3D\delta R+3DR_p},
\end{equation}
which is the same expression found for initial condition A. This makes sense since this is the case where the attractive energy of the well is dominant over the entropic potential. Therefore, the diffusive motion inside the contact cutoff does not play any role and the contact duration simply reflects the time required to overcome the energy barrier, and independent of the starting distance $R$.

\end{document}